# Artificial Intelligence based drone for early disease detection and precision pesticide management in cashew farming

Manoj Kumar R, Bala Murugan MS

School of Electronics Engineering, Vellore Institute of Technology

**Abstract**

The use of unmanned aerial vehicles (UAV) is revolutionizing the agricultural industry. Cashews are grown by approximately 70% of small and marginal farmers, and the cashew industry plays a critical role in their economic development. Several fungal and algal diseases threaten the cashew tree, resulting in substantial losses in yield. Some cashew-growing regions are particularly susceptible to powdery mildew, damping off, anthracnose, and inflorescence blight. To take timely countermeasures against plant diseases and infections, it is imperative to monitor and detect diseases as early as possible and take suitable measures. Using UAVs, such as those that are equipped with artificial intelligence, can assist farmers by providing early detection of crop diseases and precision pesticide application. To facilitate efficient and effective crop monitoring, a UAV equipped with a camera will be deployed to take aerial photographs. An edge computing paradigm of Artificial Intelligence is employed to process this image in order to make decisions with the least amount of latency possible. As a result of these decisions, the stage of infestation, the crops affected, the method of prevention of spreading the disease, and what type and amount of pesticides need to be applied can be determined. This technology has significant potential to improve the efficiency and profitability of cashew farming. UAVs equipped with sensors detect disease patterns quickly and accurately over large areas. Combined with AI algorithms, these machines can analyse data from a variety of sources such as temperature, humidity, $CO_2$ levels and soil composition. This allows them to recognize disease symptoms before they become visible. Early detection allows for more effective control strategies that can reduce costs caused by lost production due to infestations or crop failure. Using an end-to-end training architecture, mobileNetV2 determines how to classify anthracnose disease in cashew leaves. A standard PlantVillage dataset is used for performance evaluation and for standardization. Additionally, samples captured with a drone present a variety of image samples captured in a variety of conditions, which complicates the analysis. According to our analysis, we were able to identify the anthracnose with 95% accuracy and the healthy leaves with 99% accuracy.

**Introduction**

A major economic loss in the agricultural sector is caused by plant diseases, both during and after harvest. Globally, pests destroy up to 40 percent of crops each year, according to the UN Food and Agriculture Organization [11]. Controlling the spread of plant diseases at an early stage is crucial to preventing the spread of pathogens. Various historical evidence can be cited, including the Bengal famine in 1942 that caused epidemics [12]. Early detection of diseases and pathogens is essential to maintaining the sustainability of the agricultural sector. India is the second-largest producer of cereal crops like wheat and rice. Agricultural production contributes 17% of the total GDP of India [13].

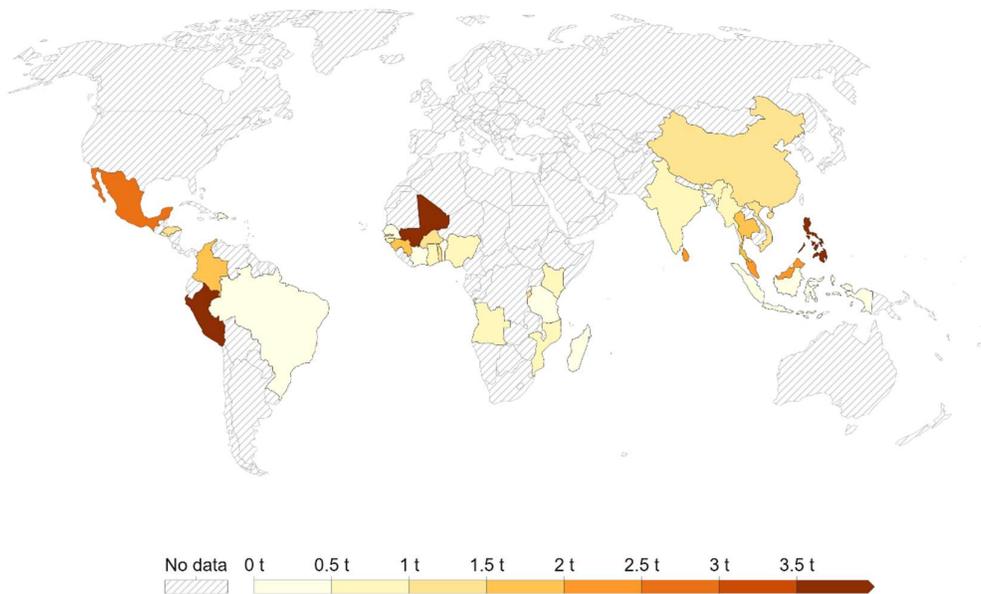

Fig. 1. Cashewnuts yield in 2020 – global scenario [10]

Furthermore, India's agriculture faces challenges from extreme weather events, including unpredictable monsoons, intense heat waves, weather anomalies, and environmental damage. Manually identifying and categorizing crop leaf diseases takes time and effort. Further complications arise from the altered size, location, and structure of the crop-diseased portion. The Tamilnadu Agricultural University states the need of crop management requires a multidisciplinary approach to handling the spreading of the disease [14, 15]. The most effective way to ensure sustainability in disease management is to use a multidisciplinary approach to disease control. Tamilnadu Agricultural University reported about 100+ disease biomarkers in cereals, millet, pulse oil seeds and cash crops. Agriculturists must take agronomically oriented measures as well as adopt technological measures such as applications of Artificial Intelligence for plant agriculture. In addition, they must adopt an Internet of Things-based approach to data aggregation and monitoring.

As a result of technological advances, image-based automated process control systems have been developed, which provide insights to agronomists and automatically identify diseased plants [16, 17]. A combination of automatic detection and timely treatment can improve crop quality and reduce epidemics by identifying diseases early, treating them promptly, and preventing their spread [41]. The affected area of a disease can be measured using image processing, and the colour difference from the unaffected area can be determined. Convolutional Neural Networks (CNNs) are known for their efficiency in pattern recognition and have been widely used for the early detection of plant diseases. It has been found that CNN has been extremely useful in identifying and classifying diseases of crop plants in recent studies [18, 19, 20]. Plant disease categorization methods extensively utilize CNN architectures such as AlexNet, GoogleNet, VGGNet, ResNet, EfficientNet, mobileNet, and Densenet.

These CNN-based deep learning models can be widely deployed in UAV-based technology to identify the disease in real time. So tracking crop growth at key stages will help provide an accurate estimate of crop yield and identify issues early. Agricultural health and soil conditions can be accurately and precisely assessed using drones equipped with infrared, multispectral, and hyperspectral sensors [21]. Using the knowledge of plant protection from agronomists, these UAVs can be operated in combination with CNN and edge computing to decide on the amount of pesticide droplets to be applied precisely in the farm after an aerial survey.

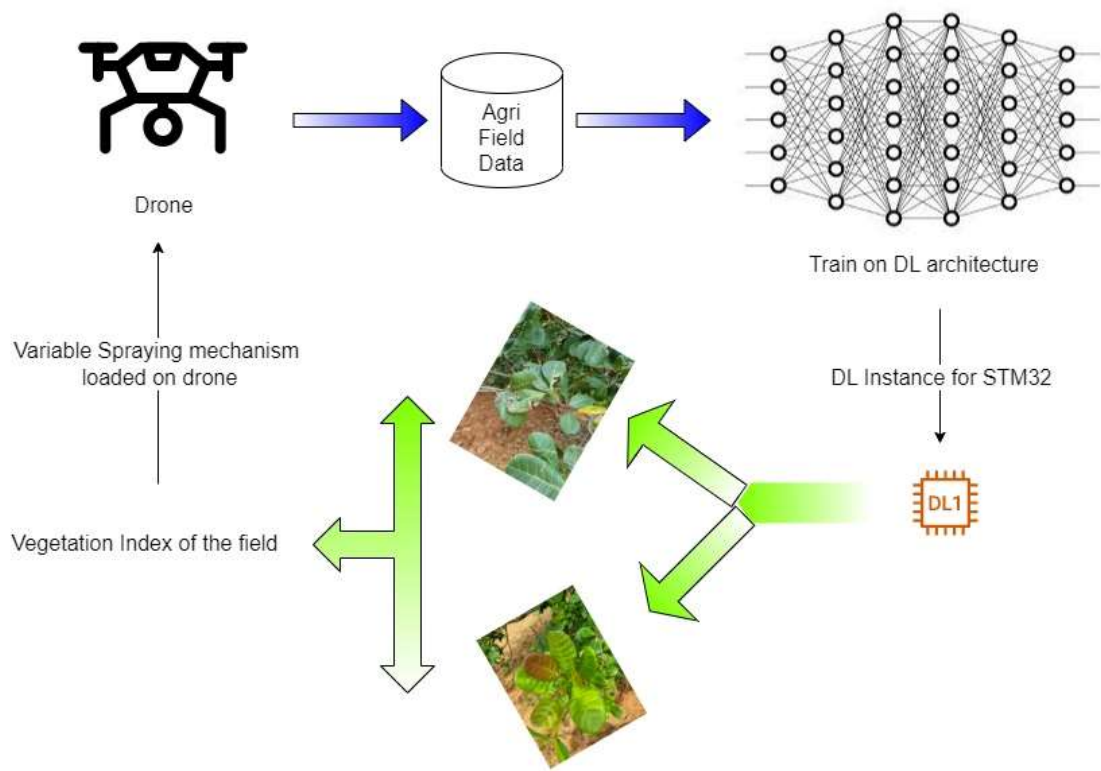

Fig. 2 Drone with variable spraying mechanism

Determining the exact disease that is spreading and affecting plants is vital. In identifying crop diseases, it is essential to understand what chemical pesticides are applied to which kind of disease to contain or kill the insecticide [22]. There is a general belief that chemical pesticides control the spread of crop diseases. A UAV-operated mechanical spraying method has become widespread among agriculturalists due to recent technological advancements that permit the application of pesticides in the field by doing a simple aerial survey and determining the amount of pesticide needed on the field [23]. In addition to its other advantages, aerial application and speed can identify and eliminate crop threats at an early stage. Many governments have taken steps to strengthen agricultural aviation, including India through its Krishi Udan 2.0 [24]. The Indian government has also formulated action plans for the implementation of Krishi Udan by adopting big technology practices in the agricultural field through public-private partnerships and through the usage of Kisan drones in the agricultural fields. Agriculturists thus benefit from adopting drone-based pesticide spraying, which facilitates targeted pesticide spraying in a more efficient manner and can produce more efficient yields. As of today, pesticide spraying is implemented by surveying the field and then spraying pesticides evenly throughout, following an aerial survey. It is proposed to modify the application of pesticides to areas where disease outbreaks are a more serious threat to plants [25]. This is while preserving pesticide application in areas that do not experience disease outbreaks. By this method, the use of pesticides can also be minimized. The adoption of this approach requires an understanding of the outbreak of disease. This can be accomplished with drones, which can capture aerial views of the field and automatically identify diseased plants. It will provide valuable insight into the agricultural field and will allow auctions to be conducted according to the recommendations of agronomists. Therefore, it is necessary to use drones that apply pesticides based on variable spraying technology rather than uniformly spraying pesticides [26]. In this way, we can also reduce the quantity of pesticides that must be applied in accordance with international standards. An understanding of how disease spreads in plants is crucial to determining the quantity of pesticides to be sprayed [27]. To analyze the situation, a drone is flown through the field and an aerial view is captured. As soon as the images are recorded with GPS-tagged information, they will have to be processed to detect the presence of diseases in crops. This can be further processed in the field through edge computing based analysis rather than the traditional cloud computing based methods where every time the data (images) has to be processed in the cloud [27].

In edge computing, computation takes place near the point where data is captured by sensors [28]. It is being done in order to enhance the instantaneous decision-making skills of the algorithm implemented in the system and, in turn, reduce the roundtrip time to send the data for computation to the cloud. In colloquial terms, edge computing refers to training an artificial intelligence model on a global scale and then executing it in a localized scenario. When compared with the Cloud, Edge Computing features location awareness of the disease affected area, mobility support in varying pesticide application methods, localization in decision making, a lower response time, contextual awareness of the disease spreading in the field, and a heterogeneous approach in varying pesticide application methods [29].

Incorporating the edge computing architecture into the drone for building an variable spraying of pesticides method involves taking instantaneous decision making skill of identifying the disease spread in the crops [30]. It will be necessary to create an artificial intelligence model in order to build this drone. An AI model will be built by fine-tuning a convolutional neural network and developing a pattern recognition algorithm for identifying diseases based on Deep Learning approaches. We have already evaluated a ResNet-based approach to identifying the disease in the Village plant dataset [31]. By augmenting data and testing in the cashew agricultural field, studies have been conducted on detecting anthracnose using an edge computing model. The drone will build by enhancing this AI based edge model to be tested in the agricultural field.

It is in this context that object detection algorithms using deep learning are developed and adopted for the purposes of localizing and classifying plant diseases. While these methods are capable of determining the exact location and type of the disease, their performance degrades when applied to a complex natural environment [32]. While deep learning architectures have been applied to a wide variety of crops and plant diseases, there is still a lot to be done to improve their robustness to generalization and accuracy when it comes to crop plant disease classification. This is especially true for novel deep-learning architectures that are applicable to crop plant disease classification.

There are numerous plant diseases that remain difficult to identify due to the fact that the affected areas have varied physical characteristics, including dimensions, contour, color, and location. To address this challenge, this study will involve building a drone-based improved mobileNet for the development of crop-disease classification tools.

In this study, we propose to make the following significant contributions:

The proposed method for building a drone to improve the performance of plant disease classification using an image classification approach could be a powerful tool in optimizing pesticide spraying. By fine-tuning this edge computing model, different pesticide application approaches can be applied in order to measure the effectiveness of the treatment. Analyzing the performance of this method will give us important insights into how accurate and efficient our drone-based system is, allowing us to make informed decisions on how best to tackle plant disease outbreaks.

**Literature Review**

Umamageshwara Rao [1] investigated the impact of aerial spraying by an unmanned aerial vehicle on rice plants. The study concludes that the impact of drone usage is not significant since the rotors in the drone have an impact on the deposition rate of droplets. Additionally, spraying pesticides at a certain height and distance also affects droplet deposition in rice and paddy fields. According to Abdul Hafeez [2], Near Infrared Cameras are being used on a daily basis to detect pests and to monitor the health of crops. Furthermore, this study investigates how drones equipped with cameras can be used to identify diseases at an early stage and how advanced image analysis tools can lead to easier decisions. Drones also assist in surveying and mapping the field precisely to determine the vegetation index. Furthermore, the study demonstrated a direct relationship between the vegetation index and harvest. A study by Tawseef and IISc, India [3] emphasizes the need for UAV-based drones to spray pesticides, particularly in populated areas like India, where 70 percent of people still rely on agriculture for income. Furthermore, the study revealed that more than a million people have been adversely affected by manual spraying of pesticides. According to Miles [4], digital technologies combined with M2M are transforming agriculture. Additionally, the study emphasizes the importance of the Internet of Things, sensors, and artificial intelligence as a crop monitoring tool in the deployment of ICT. This work also shares some results how agricultural yields can be improvised using these digital tools. According to Bian [5], machine learning and its successors can effectively increase the yield of agricultural fields. These digital technologies can also be used in India to implement precision agriculture. Furthermore, the author discusses how progressive farmers are aware of the differences in yields in the farming environment and adjust their farming practices in accordance with previous experiences.

Internationally, it is observed that AI-enabled drones are being deployed for precision farming, which directly increases crop yields. Benin, for example, exports cashews as the second largest crop. A significant amount of progress has been made in the use of artificial intelligence and drones in order to increase the yields of cashew crops in the country [6]. The government has taken the initiative to utilize these digital technologies, as cited by Annual Report of Department of agriculture, cooperation and farmers welfare. A total of 10,000 farmers have been identified as potential candidates for capacity building in the agricultural sector. Also, with the support of Technoserve, cashew farmers will implement smart agricultural practices with the assistance of the government. Additionally, the work addressed the use of smart climate practices when cultivating cashew trees. Mohapatra [7] demonstrated how smart farming practices can contribute to water conservation. This study examined how smart farming practices were implemented in Brazil by using the IoT and AI modelling tools. Also, the work highlighted how powerful data analysis can be accomplished with minimal computing power. The use of drones for aerial inspection of farm fields has been helpful to farmers in making decisions as opposed to visual inspection, according to Velusamy [8]. It is also discussed in the work how AI-based sensor technology has been adopted rapidly in the agricultural industry and how farmers were able to make informed decisions using them. In addition, the work mentions how farmers in Germany were trained to build capacity in the agricultural industries. Globally at few of the countries have started adapting to integrate digital technologies in smart farming solutions. In benin for example in partnership with Wehubit, TechnoServe is implementing the CajùLab project to promote farmer adoption of climate-smart agricultural (CSA) practices using emerging digital solutions, such as drones and machine learning [9].

[33] An airborne multispectral imaging tool has been operated by Charles Sturt University in partnership with an Australian research institute since 1994. For analyzing and tracking spatial variation in multiple agricultural crops, Charles demonstrated the effectiveness of multispectral images in terms of their qualitative content, as well as their timeliness and cost-effectiveness. Based on the results of this study, different soil zones can be sampled for soil analysis and can help locate treatment plots within paddocks. Imagery of this type was used to assess the presence of weeds early in a fallow paddock or in a seedling crop by performing a synoptic evaluation. As a result of precision farming, variations in the shape of crop emergence, as well as canopy vigor and biomass, may be induced. However, without the requisite investment or sensor equipment, it is still difficult to obtain any precision.

Daniel [34] discusses the role of precision agriculture in the modern agricultural revolution. Increasingly, farmers understand the importance of maintaining direct control over a number of critical aspects. These include the health of the crops, the irrigation system and nutrients, and other factors that may affect the crops in the field. Since the 1960s, hyperspectral imaging and multispectral imaging have been used for the analysis of agricultural crops. Despite this, there are a number of technical and practical challenges that must be overcome.

According to Tiezhu [42], the majority of agricultural research applications are focused on detecting contaminants and heavy metals, managing and assessing water sources, and monitoring crop health. In this article, the author discusses the increasing use of hyperspectral imaging in agriculture. It also highlights the challenges associated with data acquisition, processing, and analysis of hyperspectral imagery (e.g., large amounts of datasets, a high degree of dimensionality, and complexities associated with information analysis).

**Materials and Methods**

We present here the datasets and methodologies used for identifying plant diseases. Furthermore, the details of the metrics used in the evaluation are also outlined. The ultimate objective is to automatically classify healthy and diseased leaves from an image of cashew leaves. Using the drone camera, input samples including different species and diseases of the plant are collected from public datasets and from the real-world environment. To compute features and classify these images, a customized mobileNetV2 DL model is developed. It is different from previous work in that this work focuses on the classification of images in real time using a drone connected to a camera that is evaluated without internet connectivity. The following methodology is used for identifying plant diseases using MobileNetV2:

- Data collection: Collect a large dataset of plant images affected by different diseases.
- Data preprocessing: Preprocess the collected data to make it suitable for training the model.
- Model selection: Select MobileNetV2 as the deep learning model for image classification.
- Model training: Train the MobileNetV2 model on the pre-processed data using supervised learning.
- Model evaluation: Evaluate the trained model's performance on a separate validation dataset.
- Deployment: Deploy the trained model on a mobile or web-based platform for real-time predictions.
- Model fine-tuning: Finetuning the model to improve its accuracy.
- Quantization: int8 based quantization technique is used to edge optimize the model to run in the edge

This methodology can be improved based on the specific requirements and available resources.

**Dataset**

The subject area of study is located in Mudhanai, India (11.558785519004267, 79.40239239904223) which is fielding 5 acres of cashew land. The trees have an average height of 4.5 - 6m and 3 - 4.5m.

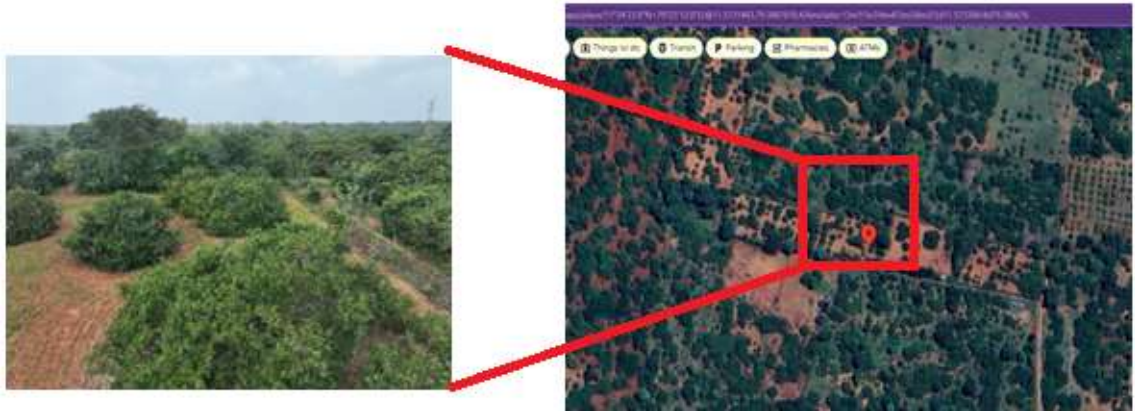

Fig. Study area in rectangle located in Mudhanai, India (11.558785519004267, 79.40239239904223)

**Methodology**

This work aimed to develop an edge computing-based architecture to distinguish between healthy and unhealthy crops. The architecture must be implemented on an unmanned aerial vehicle that is flying in the field. This work focused on using a microcontroller interfaced with a camera to perform edge computing. Microcontroller-based architectures have the advantage of minimizing the drone's battery consumption capabilities. As part of this study, we examined the capabilities of TensorFlow lite, which enables machine learning algorithms to be deployed at the edge using the platform. Fig. 2 shows a block diagram of the proposed architecture.

The proposed architecture was evaluated using the B-L475E-IOT01A ARM Cortex M4 processor running at 80 MHz to implement the proposed work. In addition to WiFi and Bluetooth capabilities, the processor is equipped with various IO ports for reading sensor data that will be necessary for future work. The microcontroller can perform edge computing and computer vision tasks required for image classification in addition to future IoT capabilities. As a key feature of this controller, it is equipped with a high-performance camera, which is necessary to collect data during the flight of the UAV. Additionally, we determined that the payload of the UAV would not be affected by the weight limitations of the identified architecture.

One of the most significant aspects of this design is the ability to identify crop diseases at the edge rather than processing the images in the cloud. This was made possible by the TensorFlow Lite model quantization feature. To ensure that the quantized model was evaluated at the edge, optimization was carried out focusing on the STM32F4 architecture. As a result, the quantized model was able to provide inference at the edge in accordance with the needs of the system.

**Experimental Validation**

The Plant Disease Detection System is designed to help farmers identify diseases that may affect their plants quickly and easily. Figure 2 illustrates the basic structure of this module.

For training and validating plant disease identification, we use the Mohanty [35] dataset of the Epidemiology lab, EPFL, Geneva. A modified mobileNet model has been deployed in the cloud for the purpose of building a smart crop disease prediction system.

The high dimensionality of image features prevents suitable visualization, even after rescaling and colour conversion. The model was evaluated and resized to 96x96 pixels. In order to comply with the microcontroller's constraints and optimize the model size, the number of samples was selected. Mobilenet V2 [36] is a convolutional neural network (CNN) architecture designed to be lightweight and efficient for mobile and embedded devices. The architecture uses depthwise separable convolutions, which reduces the number of parameters and computational costs while maintaining accuracy. The specific architecture you mentioned, Mobilenet V2 16 neuron 0.1 dropout, likely refers to a custom CNN model that utilizes the Mobilenet V2 architecture with a fully connected layer containing 16 neurons and a dropout rate of 0.1. This custom CNN model is used for crop disease identification. It is a lightweight model that can be used on a mobile device which is less computational expensive.

There are several critical points to consider when bringing a deep architecture onto a tiny processor [37, 38, 39]:

1. Model size and complexity: The model should be as small and simple as possible to reduce the memory and computational requirements of the tiny processor. This can be achieved by using lightweight architectures such as Mobilenet V2, or by pruning and quantizing the model.
2. Optimization: The model should be optimized for the tiny processor, such as by converting the model to a format that can be run efficiently on the processor's instruction set.
3. Power consumption: Tiny processors often have limited power budgets, so the model should be designed to minimize power consumption. This can be achieved by using low-power architectures, such as MobileNet V2, or by using techniques such as dynamic voltage and frequency scaling to reduce power consumption during idle periods.
4. Data pre-processing: Data pre-processing is an important step in reducing the computational requirements of the tiny processor. This can include techniques such as image resizing, data augmentation, or feature extraction.
5. Latency: The time it takes for the model to make a prediction is also important. Latency can be reduced by using techniques such as model compression and quantization, or by using techniques such as model parallelism and data parallelism to distribute the computation across multiple processors.
6. Memory constraints: Memory is a crucial constraint, especially when running deep models on tiny processors. Techniques such as model compression and quantization, or using a memory-efficient data structure can help overcome this constraint.

INT8 quantization is a technique that reduces the precision of a model's weights and activations from 32-bit floating point (FP32) to 8-bit integer (INT8) [40]. This results in a significant reduction in memory and computation requirements without significantly impacting model accuracy. TensorFlow Lite, a lightweight version of TensorFlow for mobile and embedded devices, includes support for INT8 quantization out of the box. When converting a model from FP32 to INT8, the model's weights and activations are first quantized to the INT8 range using a calibration process. This process generates a set of scale factors that are used to adjust the quantized values during inference to ensure accuracy. Once the model is quantized, it can be compressed and stored in a smaller memory space. The TensorFlow Lite's INT8 quantization feature offers a simple way to compress models from 32-bit to 8-bit, which is beneficial for memory constrained devices. This is also a standard compression technique that is widely used for mobile deployment and edge devices.

This system comprises of two major components:

1. Creating the CNN model: The first step is to create a Machine Learning model. This work is carried out using a Convolutional Neural Network (MobileNet V2) model.

a. A comprehensive study of Deep neural networks models for plan disease identification

b. Improvising the MobileNetV2 model for the dataset by deploying a "One cycle Learning Rate Policy" instead of traditional fixed learning rate

c. The hyper parameters are further optimized for weight decaying and gradient clipping to suit to the training the plant disease identification.

2. Using the model to predict: The next step is to use the created model to identify the diseases in the leaves. This step would be carried out in real-time. The model is trained on a Nvidia GPU for faster processing purposes. This model is trained using MobileNetV2 architecture for feature extraction. The model is trained for 65,89,734 params. These features are used to train class specific intermediate convolutional neural layer. This system achieved 99.35% accuracy for the dataset. The Edge computing based disease identification was tested in the agricultural field to identify the healthy and disease affected leaves and the model was able to predict the healthy leaves in realtime as shown in fig

Table 1 : The confusion matrix for the trained int8 optimized model

|  | Cashew Anthracnose | Cashew Healthy |
|---|---|---|
| Cashew Anthracnose | 95% | 5% |
| Cashew Healthy | 0.8% | 99.2% |
| F1 Score | 0.95 | 0.99 |

Table 2 : The confusion matrix for the trained float32 unoptimized model

|  | Cashew Anthracnose | Cashew Healthy |
|---|---|---|
| Cashew Anthracnose | 100% | 0% |
| Cashew Healthy | 0% | 100% |
| F1 Score | 1.00 | 1.00 |

The device performance of an INT8 vs FLOAT32 model for crop disease detection will depend on several factors, including the size and complexity of the model, the computational resources available on the device, and the nature of the validation dataset.

In general, INT8 models tend to be more computationally efficient than FLOAT32 models, as they use 8-bit integers for storage and computation, instead of 32-bit floating-point numbers. This makes INT8 models faster and less memory-intensive than FLOAT32 models, which can be particularly beneficial for real-time applications on devices with limited computational resources.

However, the accuracy of INT8 models was lower than FLOAT32 models, as INT8 models use quantization to reduce the precision of their computations. This resulted in some loss of accuracy compared to FLOAT32 models, although the extent of this loss was acceptable to suit the architecture and the validation dataset was giving an accuracy of more than 95%.

In summary, the device performance of an INT8 vs FLOAT32 model for crop disease detection depend on the trade-off between computational efficiency and accuracy, which was designed by carefully evaluating the need of specific requirements of the application.

Results

Recent experiments in the field of edge computing operated drones have yielded interesting results. Edge computing is a distributed computing system designed to streamline data processing from remote devices, like drones, by utilizing local resources instead of relying on cloud-based systems. A recent experiment employed such an approach to operate a drone with promising results. The experiment was done using an off-the-shelf drone equipped with advanced sensors, powerful processors and robust communication capabilities. It was programmed with algorithms using the edge computing system, enabling it to carry out autonomous tasks and make intelligent decisions without having to send its data over the internet or store them in the cloud. The results showed that the drone could successfully identify objects in its environment while maintaining low latency and high accuracy at all times.

The trained model and non-optimized model consumed 1.6 MB of flash memory, 89 s of onboard inference time, and 1.6 MB of RAM with 87.6% accuracy. The model was further quantized to an 8bit integer which resulted in 583.6 KB of flash memory, an inference time of 13s and 1.6MB of RAM usage. The model had an accuracy of 98% in the test data set.

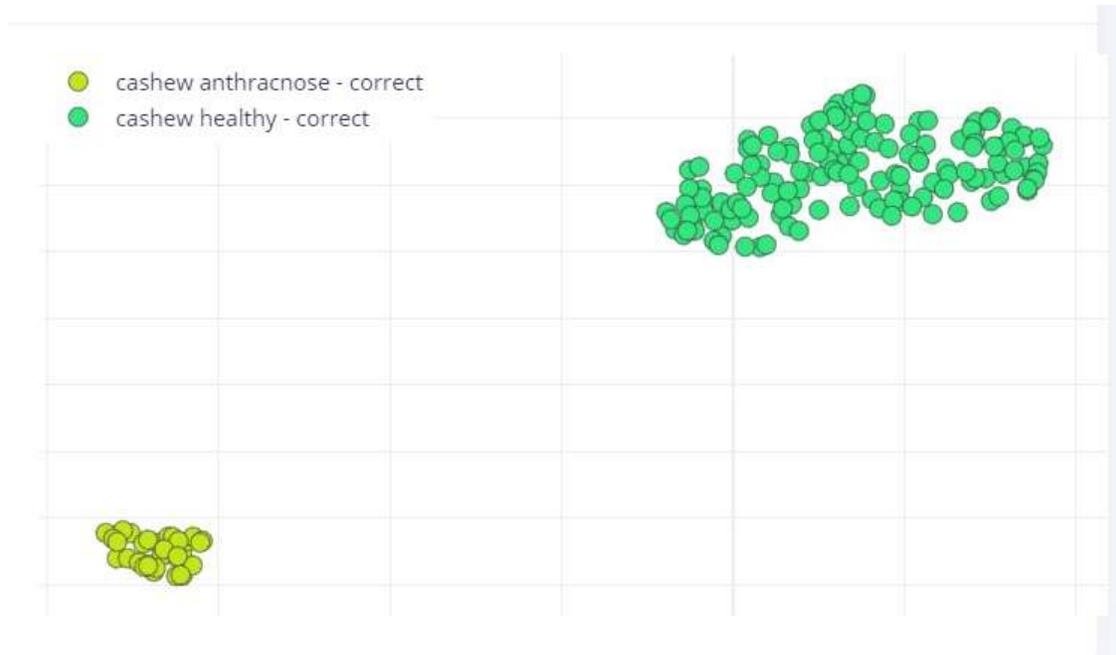

Fig. 4 Feature Explorer on cashew farming test dataset

The feature explorer graph of crop disease identification in cashew farming, visualization as in fig. 4 shows how well the model can differentiate between different features or variables in the validation dataset related to cashew farming. This visualization can help to identify the most relevant features for making accurate predictions and to determine if the model is overfitting or underfitting the data. By visualizing the model's performance, we can gain insights into the model's strengths and weaknesses, and make informed decisions about how to further improve the model's accuracy and performance.

By integrating a joint flight and mission control embedded system, we present a novel way to give drones enhanced autonomy and intelligence. With the implementation of an edge computing solution based on a powerful lightweight processing architecture and Tensorflow lite optimized inference engine, the system was able to achieve sophisticated mission objectives without significantly limiting flight duration due to a suitably designed lightweight processing architecture. The energy consumption of the system was tested in various setups and flying scenarios to demonstrate its energy efficiency. The results of the experiments were used to validate a significant sample application, and the results demonstrated the effectiveness of microcontroller integration in the payload. The system is viable in practical contexts due to its added value in terms of intelligence, which allows the system to be implemented. The proposed technology will be extended to other drone models in the future to promote broader adoption of different use cases.

Table 3: Device performance of int8 vs float 32 model on validation dataset

|  | Quantized int8 model | Unquantized float32 model |
|---|---|---|
| Inferencing Time | 970 ms | 1076 ms |
| Peak RAM usage | 130.9 K | 327.6 K |

| Flash Usage | 309.7 K | 861.6 K |

The results of using edge computing for identifying anthracnose cashew disease would depend on the specific machine learning model that was deployed on the edge device. As observed through table 3, the use of edge computing for disease identification provides faster and more efficient results compared to traditional cloud-based approaches. This is because the data was processed directly on the device, rather than being transmitted to a remote server for processing.

If more powerful edge architecture like EdgeTPU can be validated on this drone the results would have been potentially highly accurate. This would provide valuable insights into the presence and severity of the disease in cashew crops. However, it is imperative to note that the performance of the model also would be heavily dependent on factors such as the quality of the training data and the availability of resources on the edge device.

Overall, using edge computing for anthracnose cashew disease identification can provide a promising solution for addressing the challenges of disease detection in agriculture, especially in remote or resource-constrained environments as shown in fig 5.

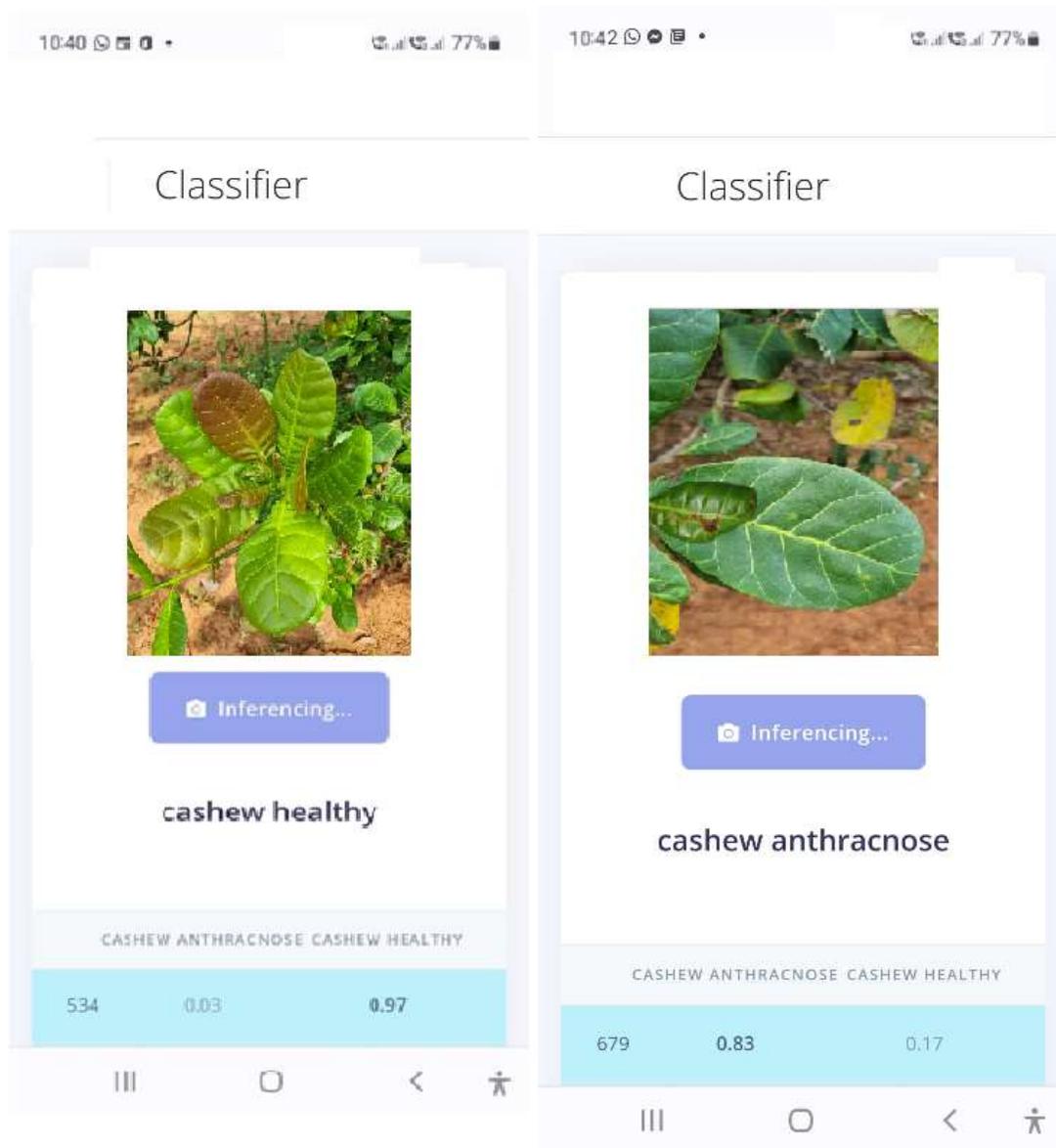

Fig. 5 Cashewnut crop healthy vs anthracnose disease classification in realtime testing carried out in field carried using a mobile connected drone

Conclusion

The role of cashew farming in India's GDP is significant. Cashew farming has provided significant economic opportunities for farmers and has contributed to the country's agricultural output, leading to increased employment and income generation. In addition, cashew farming has also been beneficial to the environment through soil conservation and improved water management.

Edge computing and drones are increasingly being used in agriculture to improve efficiency and productivity. Edge computing allows for real-time data processing and analysis at the source, rather than sending all data to a centralized location for processing. This can be particularly beneficial for large-scale agricultural operations, where data from multiple sources needs to be analyzed quickly in order to make informed decisions. Drones, on the other hand, can be used for a variety of tasks in agriculture, such as crop monitoring, soil analysis, and precision spraying. They can provide detailed, high-resolution images and data that can be used to identify issues such as crop stress, pests, and nutrient deficiencies. This information can then be used to improve crop yields and reduce the use of resources such as water and pesticides.

The original model was optimized to reduce memory usage and improve inference speed. The optimization process involved quantization to 8-bit integers, which reduced flash memory usage by 63.5%, from 1.6 MB to 583.6 KB. Inference time was also reduced by 85%, from 89 s to 13s. The RAM usage remained the same at 1.6 MB. The optimization resulted in a significant improvement in accuracy, from 87.6% to 98%. Together, edge computing and drones can provide farmers with real-time data and insights that can help them make more informed decisions, ultimately leading to increased efficiency, productivity, and reduced costs.

The use of drones for crop disease monitoring is a promising application of technology that is expected to grow in popularity in the near future. One of the key methods used for monitoring crop health with drones is the NDVI (Normalized Difference Vegetation Index) method, which involves collecting data from multispectral cameras mounted on drones.

With the NDVI method, it is possible to quickly and efficiently gather information about the health of crops across large areas, including data about the chlorophyll content of the leaves, which is an indicator of plant stress. By analyzing this data, it is possible to identify early signs of disease, monitor the spread of disease, and track the effectiveness of disease control measures.

Drones equipped with NDVI cameras can fly over crops and collect data without disturbing the plants or soil, making it an ideal solution for crop disease monitoring in remote or hard-to-reach areas. The data collected by the drones can be transmitted to a remote server for analysis, providing farmers and agronomists with a real-time view of the health of their crops.

In future work, the use of drones for crop disease monitoring using the NDVI method is a highly innovative solution that has the potential to revolutionize the way that crop health is monitored and managed. By providing fast, efficient, and non-invasive data collection, this technology has the potential to help farmers and agronomists to protect their crops and improve yields.